\documentstyle[11pt]{article}
\begin{document}
\begin{center}

{\bf A search for spectral variations of planetary  nebulae and related objects}\\
\vspace{8mm}
 \large
 L.N. KONDRATYEVA

 \vspace{4mm}

{\it Fessenkov Astrophysical Institute, 480020, Almaty, Kazakhstan}\\

    \vspace{4mm}

\end{center}
\vspace{4mm} \small \noindent The results of long-term spectral
observations  were used to search for changes in planetary nebulae
and emission-line stars. Significant increase of excitation degree
is found in two objects: M1-6 and M1-11.\\
\vspace{6ex} \noindent{KEY WORDS\,}\,\rm\mbox{planetary  nebulae,
-- spectra: variability}

     \normalsize
\section{INTRODUCTION }
It is widely known, that physical parameters of planetary nebulae
and their central stars change gradually during evolution, but
these effects are rather difficult for detection in a short time
interval Sometimes much more
 exotic processes, such as throwing out of a secondary envelope, jets and
 so on, take place in planetary nebulae.   Similar processes cause
 appreciable changes in spectra, but   these events are rare and unpredicted.
In other words everyone knows, that planetary nebulae are
changing, but these effects are hard to be found out. The problem
is more complicated, because results, obtained simultaneously, but
by different authors may sometimes differ sufficiently.  We
decided to use the fact, that spectral observations of the large
group of planetary nebulae and other emission  objects were
carried out in the Astrophysical Institute for a long time and
with the same equipment. The very first spectrograms have been
obtained in 1970 - 1973, and the last observations  were conducted
in 2004 - 2005.
      The aim of this paper is to search for any variations in spectra
of chosen objects within 30 - 35 years. We expected, that even
small fluctuations of  T$_{eff}$ ,  T$_e$ or N$_e$\rm\, in some
young objects can affect essentially the  intensities of forbidden
lines.  We also hoped, to find any traces of fast dynamic
processes, such a jet,  if they have taken place within the
specified period.

\section{The sample of objects}
The objects of different types and ages were chosen for this
study. There are three classical planetary nebulae of high and
moderate excitation (K3-58, K3-3, M1-5), three low -excitation
planetary nebulae (M1-6, M1-11, M1-12) and M1-77. The last object
has the spectrum of very low-excitation with the [O\,{\sc ii}],
[N\,{\sc ii}] and [S\,{\sc ii}] emission lines, weak emission and
absorption lines of neutral metals. It was identified as a young
planetary nebula by \cite{sab83}.  Later some irregular changes of
brightness and radial
 velocity  were found out by \cite{han99} and \cite{dem04}. Our observations showed that
  there are  TiO  bands   in the spectrum of M1-77, and it  was considered to be the
   nebula around the symbiotic star \cite{kon05}. In addition three emission objects
    were included in our program:  MWC 137, M1-15 and He2-446.

\section{Observations}
 All observations have been carried out
by the author with the 70--cm telescope of the Astrophysical
Institute (Almaty, Kazakhstan). We used a slit spectrograph,
equipped with a three-cascade image tube until 1998, and with a
CCD matrix ST8 since 1999. A sample of gratings and objective
lenses provided a spectral range from 3700 to 8200 \r{A}, with a
dispersion in the range of 20 -- 200 \r{A}/mm and with 0.3 -- 1.1
\r{A}/pixel after 1998. Wavelength calibration has been done using
a laboratory source of He\,{\sc i}, Ne and Ar emission lines.

The earlier  data  base has been obtained in 1970 - 1975, and the
most of results (but not all) were published. Later,  observations
of  objects were continued but quite
 irregularly - from time to time until 2005.

\section{Observational data for the planetary nebulae}
At the first stage of our study we compared the results of the
earliest and the latest observations. One or two rows of data for
the middle of specified period were included as well (Table 1).
The ratios of I(O\,{\sc iii},4959+5007) / I(H$_{\beta}$)\rm\, and
I(N\,{\sc ii},6548+6583)/I(H$_{\alpha}$)\rm\, were used, as an
initial criterion. These lines were chosen because, as a rule,
they are the strongest in a spectrum, and secondly, an influence
of interstellar absorption on these ratios  is  minimal. The
designation of the objects, observation date  and ratios of line
intensities are given in the 1 - 4 columns of Table 1.
Corresponding references are presented in the last column. Our
unpublished results are marked by "AFI".

One can see, that line ratios for the most of the objects in Table
1 do not show significant changes, i.e. ratios of the maximal
values to minimal do not exceed 1.05. By the way this result
confirms the reliability of our method of observations and
measuring. But spectra of two objects (M1-6 and M1-11) have
undergone essential changes: I(O\,{\sc iii},4959+5007) /
I(H$_{\beta}$)\rm\, increased during the specified period by 40\%
and 110\% , respectively.  A little bit smaller increasing of this
ratio ( about 14\%)  is noticed in the spectrum of K3-58. Below we
will consider the spectral changes of M1-6 and M1-11 in details.
In the Table 2, 3 results of spectral observations for these
objects are presented. The year of observation is given in the
first line, The next line contains references.   All values of
emission line intensities are dereddened.

Six low lines contain available parameters of nebulae and central
stars: EW -- the equivalent widths, log(F$_{\beta}$)--logarithm of
the absolute flux at H$_{\beta}$ (in erg cm$^{-2}$
sec$^{-1}$)\rm\, N$_e$ -- the electron density determined from the
[S\,{\sc ii}],6717/6731 line ratio, T$_e$-- the electron
temperature calculated from [N\,{\sc ii}]line ratio and/or
[O\,{\sc iii}]line ratio and T$_{eff}$\rm\, determined by Zanstra
method.Analysis  of  Table2 shows that in M1-6 the first essential
increase of
 [O\,{\sc iii}]  intensities have taken place between 1975 and 1986, and at  that time
changes did not concern other zones of the nebula. The
strengthening of He\,{\sc i} lines  began rather later, and
decrease of [S\,{\sc ii}] was registered  only in 2004. All these
facts give evidence of gradual increase of excitation degree of
nebula. But we may say nothing about its reason. Both the
temperature of central star T$_{eff}$\rm, and electron  density
N$_e$ in S$^+ $\rm\, region look like rather constant. Some other
parameters, such as EW(H$_{\beta}$)\rm\, and F(H$_{\beta}$)\rm\,
show fluctuations, but without any tendency. It is appeared that
an emission spectrum is more sensitive
 to changing  of conditions  in a nebula than usual methods of parameters
estimation.

     Spectral data for M1-11 are presented in Table 3. In spite of  low
precision there is good agreement
 between our the earliest values of [O\,{\sc iii}] intensities,
 and data of  \cite{sha89}. Then some weakening  of [O\,{\sc iii}]
intensities was registered,  and errors were decreased
significantly. Intensity of [O\,{\sc iii}] lines began to increase
gradually since  1988. Up to now other lines do not show essential
change.  Only the intensity of He\,{\sc i}  has a little raised.
In other word the raising  of excitation degree takes place only
in the most inner region of nebula M1-11.
 Some physical parameters of M1-11 have  undergone fluctuations.
Increasing  of T$_e$(O\,{\sc iii})\rm\,  as high as twice after
1986  is very interesting fact. It may be the result of any
dynamic process (shock wave?) in the inner zone of the nebula.
From the other hand high value of T$_e$\rm\, may be the reason of
strengthening of forbidden lines. The value of
EW(H$_{\beta}$)\rm\,  is practically constant, but EW(5007)
increases in according to  the intensity behavior.
\section{Spectra of related objects}
 Three other objects from our list are  identified  as B[e] stars. Their spectra
contain the strong continuum and Balmer emission lines of
hydrogen.  In addition there are [O\,{\sc i}] 6300,6364\r{A} lines
in the spectrum of He2-446 and weak emission lines of Fe\,{\sc ii}
in the spectrum of M1-15 and MWC 137. All three  objects are
embedded into large H\,{\sc ii} regions and have infrared excesses
due to warm dust. Emission lines H$_{\alpha}$\rm\,
 and
H$_{\beta}$\rm\, are broad in all three objects.

 Filter
photometric and  high resolution spectral observations have been
carried out in order to
 study the  stellar brightness and profiles of emission lines.
  The corresponding procedure  is  described in details
 in \cite{kon03}.

 Selective data for these  objects are presented in Table 4.  This information,
   is not complete, but it
allows to estimate the behavior of studied stars.\\
 \noindent{\bf{He2-446}\rm\,} is classified as a B[e] star \cite{win01}. It shows a
photometric variability within $0^m2$\rm.  There is no correlation
between change of brightness and  EW values. One can see that
between 1988 and 1995 EW(H$_{\alpha}$)\rm\, and
EW(H$_{\beta}$)\rm\, have increased as high as twice. At the same
time FWHM has remained practically unchanged. The profile of
H$_{\alpha}$\rm\, consists of two components with the following
ratio of their intensities: I$_{blue}$/I$_{red}$ =0.70$\pm
$0.05\rm\,. It may be explained as a result  of stellar wind and
rotating. Variations of EW is probably connected with a gradual
increase of ionized mass of stellar
envelope.\\
 \noindent {\bf{M1-15}\rm\,} is the faintest among the chosen stars,
and thus it is  poorly studied. All that we know are the values of
EW(H$_{\alpha}$)\rm\,,
and their  changes are within errors.\\
\noindent {\bf{MWC 137}\rm\,} -- the central star of H\,{\sc ii}
region Sh2-266. According  to \cite{est98} it is an evolved  B[e]
supergiant, probably of B0 spectral class. As it is seen from
Table 4, EW(H$_{\alpha}$)\rm\, and EW(H$_{\beta}$)\rm\, show
significant changes, as far as V-magnitude  Spectral and
photometric variability of the object may be coursed by change of
stellar wind power.

\section{Conclusions}Long-term spectral observations of planetary nebulae and
some emission-line stars were carried out within 30-35 year. The
results were analyzed in order to find out  any change in these
objects. Fluctuations of  some emission line intensities in the
most of objects did not exceed 5\%. Essential changes were
revealed in two planetary nebulae: M1-6 and M1-11. Significant
strengthening of [O\,{\sc iii}] and He\,{\sc i} lines and
weakening of [S\,{\sc ii}] (in M1- 6) testifies about change of
physical conditions in the nebulae. It is reasonable to assume
that all observable effects are caused by increase of
T$_{eff}$\rm\,, however available estimations of this stellar
parameter don't confirm such conclusion. In case of M1-11 a sudden
increase of N$_e$\rm(S\,{\sc ii}), T$_e$(O\,{\sc iii})\rm\, and
T$_e$(NII)\rm\, was registered in 1996. Probably these changes are
connected with some dynamic events in the nebula
 Our study of  the emission-line stars shows, that MWC137 has undergone as
photometric and spectral variability. All changes are irregular
and have rather low amplitude.  M1-15 is appeared to be quite
stable object. He2-446 showed significant increase of EW of
hydrogen lines, connected probably with a growth of ionized mass
of circumstellar envelope.

\begin{table}

\begin{center}
Table 1. Spectral data for the chosen planetary nebulae
\end{center}
\begin{center}
\begin{tabular}{ccccc}
\hline
Name  & Date & I[O\,{\sc iii}]/I(H$_{\beta}$) &I[N\,{\sc ii}]/I(H$_{\alpha}$)& References\\
 \hline
& 1974-1975 & 14.5$\pm$ 2.0& 1.73$\pm$ 0.24& AFI \\
K3-58& 1988& 15.2$\pm$ 1.5& 1.62$\pm$ 0.16& \cite{ack92} \\
& 2004-2005 & 13.3$\pm$ 1.0& 1.59$\pm$ 0.10& AFI \\

\hline
& 1972-1973 & 4.82$\pm$ 0.48& 0.44$\pm$ 0.05& \cite{bar78} \\
M1-5& 1971-1973 & 5.02$\pm$ 0.50& 0.43$\pm$ 0.04& \cite{kon78} \\
& 2004-2005 & 5.08$\pm$ 0.45& 0.44$\pm$ 0.04& AFI \\
\hline
& 1971-1973 & 1.25$\pm$ 0.12& 0.60$\pm$ 0.07& \cite{kon79} \\
M1-6& 1986 & 1.60$\pm$ 0.16& 0.68$\pm$ 0.07& \cite{ack92} \\
& 2004-2005 & 1.72$\pm$ 0.11& 0.76$\pm$ 0.04& AFI \\
\hline
& 1986 & 0.10$\pm$ 0.02& 0.88$\pm$ 0.10& \cite{kal96} \\
M1-11& 1996 & 0.14$\pm$ 0.01& 0.98$\pm$ 0.10& \cite{dop97} \\
& 2004-2005 & 0.21$\pm$ 0.04& 0.95$\pm$ 0.05& AFI \\
\hline
& 1975 & 0.26$\pm$ 0.06& 0.74$\pm$ 0.06& \cite{kon79}\\
M1-12& 1993 & 0.25$\pm$ 0.04&               & \cite{cui96} \\
& 2004 & 0.25$\pm$ 0.03& 0.72$\pm$ 0.04&  AFI\\
\hline
& 1972-1973 &               & 0.37$\pm$ 0.05& AFI \\
M1-77& 1983 &               & 0.33$\pm$ 0.06& \cite{sab83} \\
&1988       &               &0.41$\pm$ 0.05& \cite{ack92} \\
& 2004-2005 &               & 0.41$\pm$ 0.04&  AFI\\
\hline
& 1972-1974 &11.8 $\pm$ 1.2 & 4.45$\pm$ 0.60&  AFI\\
M3-3 & 1987 &11.6 $\pm$ 0.9 & 4.36$\pm$ 0.40& \cite{pei87} \\
& 2005      &12.1 $\pm$ 0.8 & 4.40$\pm$ 0.40& AFI \\
\hline

\end{tabular}
\end{center}
\end{table}

\begin{table}[t]

\begin{center}
Table 2. Results of spectral observations of M1-6
\end{center}
\begin{center}
\begin{tabular}{ccccccc}
\hline

Date of     & 1971-1975   &1986 & 1989-1990&    1995    & 1996       &2004-2005\\
observations&     &     &          &            &                  \\
\hline
References  & \cite{kon79}      &\cite{ack92}&AFI&AFI&\cite{dop97}&AFI\\

\hline


 4101 H$_{\delta}$ & 23.6$\pm$3.0&            &25$\pm$4  &            &            &25.2$\pm$2.5\\
 4340 H$_{\gamma}$ & 47$\pm$3.0  &            &48$\pm$3  & 47.5$\pm$5 &            &44.0$\pm$3.2\\
 4471 He\,{\sc i}          & 7.9 $\pm$1.0&            &          &            &            &10.1$\pm$1.5\\
 4861 H$_{\beta}$  & 100$\pm$3 &100            &100$\pm$3 &100$\pm$4  &100$\pm$2   &100$\pm$3 \\
 4959 [O\,{\sc iii}]       & 29.0$\pm$3.0 &            &42$\pm$4 & 39.4$\pm$4 &            &44.1$\pm$2.3\\
 5007 [O\,{\sc iii}]       & 86.7$\pm$7.7 &122         &117$\pm$7 &122$\pm$6  &            &128$\pm$6.7\\
 6548 [N\,{\sc ii}]        & 47.2$\pm$3.5 &         & 46.4$\pm$4.0&            & 45.8$\pm$4.1&48.5$\pm$3.5\\
 6563 H$_{\alpha}$ & 285$\pm$15   &  285       & 285$\pm$7&            & 286$\pm$20&286$\pm$11 \\
 6583 [N\,{\sc ii}]        & 124$\pm$10   &  140       &130$\pm$11&            & 135.6$\pm$11.1&146$\pm$8 \\
 6678 He\,{\sc i}          & 2.1$\pm$0.5  &            &2.3$\pm$0.5&            &            &4.2$\pm$0.4  \\
 6717 [S\,{\sc ii}]        & 5.4$\pm$0.5  &  1.11      &4.8$\pm$0.6&            & 5.04$\pm$0.5&2.9$\pm$0.3 \\
 6731 [S\,{\sc ii}]        & 7.9$\pm$0.7  &  2.46     &8.4$\pm$0.9 &            & 7.16$\pm$0.7&4.2$\pm$0.4\\
 EW(H$_{\beta}$)  & 260$\pm$60  &            &187$\pm$23&            &  257$\pm$20&230$\pm$20\\
 EW(5007)         &             &            &213$\pm$23&            &            &260$\pm$30 \\
 log F(H$_{\beta}$) &-12.20$\pm$0.10&       &            &            & -12.269$\pm$0.005 &   \\
 N$_e$[S\,{\sc ii}]        &8300$\pm$1000 &        &8000$\pm$1000&           &  7500$\pm$1000  & 8000$\pm$1000\\
 T$_e$[N\,{\sc ii}]        &12000$\pm$1000&        &             &           &                 &           \\
 T$_{eff}$         &$\ge$30000&        &             &           &                 & $\ge$29300 \\
\hline

\end{tabular}
\end{center}

\end{table}

\begin{table}

\begin{center}
Table 3. Results of spectral observations of M1-11
\end{center}
\begin{center}

\begin{tabular}{lccccccc}
 \hline
Date of                  & 1971-   &1984             &1986           & 1988-  &    1996         & 1999       &2004-\\
observ.             &  1973   &                 &               &  1989  &                 &            & 2005        \\
\hline
Refer.               & \cite{kon79}&\cite{sha89}     &\cite{kal96}   &AFI         &\cite{dop97}      &AFI         &AFI\\

\hline

\hline
 4101 H$_{\delta}$& 24.2$\pm$3.2  &                 & 23.1$\pm$0.5  &             &                 &25.8$\pm$2.7  &26.4$\pm$1.7\\
 4340 H$_{\gamma}$& 51$\pm$5.2    &                 & 43$\pm$0.6    &             &                 &45.5$\pm$4.9  &44.0$\pm$3.3\\
 4861 H$_{\beta}$& 100$\pm$3     &100              &100$\pm$4      &100$\pm$2    &100$\pm$3        &100$\pm$4     &100$\pm$3   \\
 4959 [O\,{\sc iii}]&  6.0$\pm$2              &5.2$\pm$2      &2.2$\pm$0.1  &3.4$\pm$0.5      &3.3$\pm$0.04  &5.5$\pm$0.7   &5.3$\pm$0.3\\
 5007 [O\,{\sc iii}]& 19  $\pm$8              &               &6.8$\pm$0.2  &10.2$\pm$1.4     &10.4$\pm$0.1  &15.4$\pm$1.9  &15.6$\pm$0.7\\
 6548 [N\,{\sc ii}]& 62.5$\pm$5.5            &               &63.7$\pm$2.2 & 65$\pm$6        &66$\pm$4      &69.5$\pm$8    &65.5$\pm$3.5\\
 6563 H$_{\alpha}$& 285$\pm$15    &285$\pm$7        & 285$\pm$10    &286$\pm$14   &285$\pm$10       &286$\pm$20    &285$\pm$10 \\
 6583 [N\,{\sc ii}]& 186$\pm$10    &192$\pm$5        & 187$\pm$10    &190$\pm$12   &201$\pm$10       &200$\pm$20    &204$\pm$10 \\
 6678 HE\,{\sc i}& 0.8$\pm$0.4   &                 &0.85$\pm$0.04  &1.05$\pm$0.10&1.19$\pm$0.05    &1.1$\pm$0.2   &1.3$\pm$0.1\\
 6717 [S\,{\sc ii}]& 0.42$\pm$0.08 &                 &0.63$\pm$0.04  &0.70$\pm$0.09&0.46$\pm$0.03    &0.46$\pm$0.03 &0.51$\pm$0.05\\
 6731 [S\,{\sc ii}]& 0.96$\pm$0.15 &                 &1.1$\pm$0.04   &175$\pm$30   &0.96$\pm$0.05    &0.95$\pm$0.11 &1.06$\pm$0.08\\
 EW(H$_{\beta}$)        &               &                 &               &18$\pm$1.4   &173$\pm$17       &175$\pm$20    &175$\pm$18\\
 EW(5007)               &               &                 &               &             &19$\pm$2         &40$\pm$10     &30$\pm$7\\
 logF(H$_{\beta}$)      &-11.76         &-11.839          &11.85          &             &-11.781          &               &          \\
                        &$\pm$0.10      &$\pm$0.005       &$\pm$0.05      &             & $\pm$0.003      &               &           \\

 N$_e$[S\,{\sc ii}]     &10000          &                 &4500           &5100         &9300             & 8300          &8300      \\
                        &$\pm$2000      &                 &$\pm$500       &$\pm$900     &$\pm$900         &$\pm$900       &$\pm$800\\

 T$_e$[N\,{\sc ii}]     &10500          & 9800            &10300          &             &22800            &              &             \\
                        &$\pm$1000      &$\pm$1000        &$\pm$400       &             &  $\pm$800       &              &             \\
 T$_e$[O\,{\sc iii}]    &               &                 &11300          &             &22000            &              &           \\
                        &               &                 & $\pm$400      &             &$\pm$800         &              &             \\
 T$_{eff}$              &$\ge$26500     &29000            &27000          &             &                 &              &$\ge$27000  \\
\hline
\end{tabular}

\end{center}

\end{table}

\begin{table}[t]

\begin{center}
Table 4. Spectral data for the chosen B[e] stars
\end{center}
\begin{center}
\begin{tabular}{ccccccc}
\hline
Name  & Date of    & EW(H$_{\beta}$) & EW(H$_{\alpha}$        &FWHM(H$_{\alpha}$)& Vmag&Refrences\\
      &observations& in \r{A}\rm     &  in \r{A}\rm           & in \r{A}\rm      &\\
 \hline
& 1971-1973        & 36$\pm$5   & 280 $\pm$30   &6.6$\pm$0.4&14.$^m$7$\pm$0.03&AFI \\
& 1988             & 35$\pm$4   & 330$\pm$30    &           &14.$^m$6$\pm$0.2&AFI\\
He2-446&1995-1996  & 70$\pm$15  & 700$\pm$40    &           &            &AFI\\
& 2004-2005        & 68$\pm$7   & 730$\pm$60    &           &            &AFI \\
\hline
& 1971-1973        &            & 2453$\pm$30   &7.8$\pm$0.4&            &AFI \\
M1-15&1991         &            &230$\pm$17     &           &            &AFI \\
& 2004-2005        &            &245$\pm$12    &           &            &AFI \\

\hline
& 1971-1973        & 47$\pm$10  & 300$\pm$40    & 5.5$\pm$0.4&            &AFI\\
& 1977-1980        & 46$\pm$10  &               &            & 11.$^m$67-11.$^m$98 &\cite{win01} \\
& 1981             &            &311$\pm$15     &            &             &AFI\\
& 1988             & 54.7$\pm$10&240$\pm$25     & 4.8$\pm$0.4&             &AFI\\
MWC137&1989-1994   &            &254$\pm$27     & 4.6$\pm$0.4&             &\cite{est98}   \\
&1993              & 60$\pm$10  &295$\pm$10     &            &             &AFI\\
&1996              &            &550$\pm$10     &            & 11.$^m$2        &\cite{oud99}\\
& 1998             & 39$\pm$10  &200$\pm$20     &            &             &AFI\\
& 2004-2005        & 37$\pm$5   &196$\pm$10     & 5.0$\pm$0.3  &             &AFI \\
\hline
\end{tabular}
\end{center}
\end{table}

\begin{thebibliography}{17}
\bibitem{sab83}
F. Sabbadin, et al., Astron\&Astrophys.\bf{123}\rm\, 147 (1983).
\bibitem{han99}
G. Handler. Astron\&Astrophys.\bf{135}\rm\, 493 (1999).
\bibitem{dem04}
O. De Marco, et al., Astrophys. J. \bf{602L}\rm\, 93(2004).
\bibitem{kon05}
L. N. Kondratyeva.  Izvestiya of AFIF. \bf{4}\rm\, 100 (2005).
\bibitem{kon78}
L. N. Kondratyeva. Astron. Zh. \bf{55}\rm\, 334 (1978).
\bibitem{ack92}
A. Acker. Catalogue of Galactic Planetary nebulae. (1992).
\bibitem{bar78}
T. Barker.  Astrophys. J. \bf{219}\rm\, 914 (1978).
\bibitem{kon79}
L. N. Kondratyeva. Astron. Zh. \bf{56}\rm\, 345 (1979).
\bibitem{kal96}
J. Kaler, K. Kwitter. Publ. Astron. Soc. Pacif. \bf{108}\rm\, 980
 (1996).
\bibitem{dop97}
M. Dopita, C. Hua. Astrophys. J.Supple Ser. \bf{108}\rm\, 515
 (1997).
\bibitem{pei87}
M. Peimbert, S. Torres-Peimbert.
 Rev. Mex. Astron Astrof. \bf{14}\rm\, 540 (1987).
\bibitem{cui96}
F. Cuisiner, A. Acker,  and J. Koppen. SAstron\&Astrophys.
 \bf{307}\rm\, 215 (1996).
\bibitem{sha89}
R. Shaw, J. Kaler. Astrophys. J. Supple Ser. \bf{69}\rm\, 495
 (1989).
\bibitem{est98}
C. Esteban, M. Fernandez. Mon. Not. R. Astron. Soc. \bf{298}\rm\,
185 (1998).
\bibitem{oud99}
R. Oudmajjer, J. Drew.  Mon. Not. R. Astron. Soc. \bf{305}\rm\,
166 (1999).
\bibitem{win01}
D. de Winter, M. Ancker, et al., Astron\&Astrophys. \bf{380}\rm\,
609 (2001).
\bibitem{kon03}
L. N. Kondratyeva.  Astron\&Astrophys. Transactions.
 \bf{22}\rm\, 181 (2003).


\end{thebibliography}
\end{document}